\documentstyle[twoside,fleqn,espcrc2,psfig]{article}

\newcommand{\AmS}{{\protect\the\textfont2
  A\kern-.1667em\lower.5ex\hbox{M}\kern-.125emS}}

\newcommand{\be}{\begin{equation}}
\newcommand{\ee}{\end{equation}}
\newcommand{\ben}{\begin{eqnarray}}
\newcommand{\een}{\end{eqnarray}}

\newcommand{\chiral}{\bar{\psi} {\psi}}
\newcommand{\<}{\langle}
\renewcommand{\>}{\rangle}
\newcommand{\Tr}{\mbox{\textrm{Tr}}}

\title{Screening mass responses to chemical potential at 
finite temperature\thanks{Talk presented by Y.~Liu}}

\author{QCD-TARO Collaboration:
	S.~Choe${}^a$,
        Ph.~de~Forcrand${}^b$,
        M.~Garc\'{\i}a~P\'erez${}^c$,
        S.~ Hioki${}^d$,
        Y.~Liu$^{\rm a}$, 
        H.~Matsufuru${}^e$,
        O.~Miyamura$^{\rm a}$,
        A.~Nakamura${}^f$,
        I.-O.~Stamatescu${}^{g,h}$,
        T.~Takaishi${}^i$,
        and 
        T.~Umeda${}^j$\\
\vspace{4mm}
${}^a$ Department of Physics, Hiroshima University, 
       Higashi-Hiroshima 739-8526, Japan\\
${}^b$ Institut f\"ur Theoretische Physik,
       ETH-H\"onggerberg,CH-8093 Z\"urich,Switzerland \\
${}^c$ Theory Division, CERN, CH-1211 Geneva 23, Switzerland \\
${}^d$ Department of Physics, Tezukayama University,
       Nara 631-8501, Japan \\
${}^e$ Yukawa Institute for Theoretical Physics,
       Kyoto University,  Kyoto 606-8502, Japan \\
${}^f$ IMC, Hiroshima University,Higashi-Hiroshima 739-8521, Japan \\
${}^g$ Institut f\"ur Theoretische Physik, Universit\"at 
       Heidelberg,  D-69120 Heidelberg, Germany \\
${}^h$ FEST, Schmeilweg 5, D-69118 Heidelberg, Germany \\
${}^i$ Hiroshima University of Economics, Hiroshima 731-0192, Japan \\
${}^j$ Center for Computational Physics, University of Tsukuba, 
       Tsukuba 305-8577, Japan \\
}

\begin{document}

\begin{abstract}
Responses to chemical potential of the pseudoscalar meson screening 
mass and the chiral condensate in lattice QCD are investigated. 
On a $16 \times 8^2 \!\times\! 4$ lattice with two flavors of staggered quarks 
the first and second responses below and above $T_c$ are evaluated. 
Different behavior in the low and the high temperature phases 
are observed, which may be explained as a consequence of the chiral 
symmetry breaking and restoration.  
\end{abstract}
\maketitle

\section{Introduction}
\vspace{-3mm}
The study of finite baryon density at finite temperature is not only very 
useful to  understand the phase transition between hadron and quark
gluon plasma, but also quite important for heavy ion collision experiments 
which require theoretical understanding of hadronic properties at 
finite baryon density and temperature~\cite{NA50}.   
 As the fermionic determinant at finite chemical potential 
is complex 
on the lattice, numerical simulations are very difficult    
but the quenched approximation at finite chemical 
potential can give physically misleading results~\cite{Stephanov}; 
simulations with dynamical fermions are therefore essential to extract the relevant 
physics~\cite{Hatsuda-Lee}.     

We propose a new technique to investigate non-zero chemical potential 
using lattice QCD simulations. 
A Taylor expansion in $\mu$ is used in the vicinity of zero 
$\mu$ at finite temperature for evaluating masses and chiral condensate.
Here we obtain the Taylor coefficients directly in a simulation
at $\mu=0$, by measuring the derivatives of the relevant observables. 
There is in fact much interesting physical information which 
can be extracted from the behavior of a system at small chemical potential.
Our preliminary results have been reported in~\cite{TARO1}.

\vspace{-3mm}
\section{Lattice formulation}
\vspace{-3mm}
Assume that the spatial hadron correlator C(x) is dominated by a single 
pole contribution,
\ben
 C(x)
  &\equiv&  \sum_{y,z,t}\<H(x,y,z,t)H(0,0,0,0)^{\dagger}\>
   \nonumber \\
  &=& A ( e^{-\hat{M}\hat{x}} +e^{-\hat{M}(L_x-\hat{x})} ) ,
\label{singlepole}
\een
where $\hat{M}=aM$ and $\hat{x}=x/a$. $L_x$ is the lattice size in the 
$x$-direction. 

We take the first and second derivatives of the hadron correlator
with respect to $\hat{\mu}\equiv a \mu = \mu/(N_t T)$,
where $\mu$ is the chemical potential.
\ben
\label{deri-1}
\frac{1}{C(x)} \frac{dC(x)}{d\hat{\mu}}
 &=& \frac{1}{A} \frac{d A}{d\hat{\mu}}
 \\ \nonumber
 & & \hspace{-3.0cm}
 + \frac{d\hat{M}}{d\hat{\mu}}
  \left\{ \left( \hat{x}-\frac{L_x}{2} \right) \tanh
     \left[ \hat{M} \left( \hat{x}-\frac{L_x}{2} \right) \right]
    - \frac{L_x}{2} \right\} ,
\een
and
\ben
\label{deri-2}
\frac{1}{C(x)} \frac{d^2 C(x)}{d \hat{\mu}^2}
 &=&  \frac{1}{A} \frac{d^2 A}{d\hat{\mu}^2} \\ \nonumber
& & \hspace{-3.2cm}
     + \left( \frac{2}{A} \frac{d A}{d\hat{\mu}} 
          \frac{d \hat{M}}{d\hat{\mu}} +
         \frac{d^2 \hat{M}}{d\hat{\mu}^2} \right) \\ \nonumber
& & \hspace{-3.2cm}
      \left\{ \left( \hat{x}-\frac{L_x}{2} \right) \tanh
             \left[ \hat{M} \left( \hat{x} - \frac{L_x}{2} \right)
         \right]  - \frac{L_x}{2} \right\} \\ \nonumber
& & \hspace{-3.2cm}
+\left( \frac{d \hat{M}}{d\hat{\mu}} \right)^2
   \left\{ \;\;
       \left( \hat{x}-\frac{L_x}{2} \right)^2 
       + \frac{L_x^2}{4} \right.
\nonumber \\ \nonumber
& & \left. \hspace{-1.9cm} 
- L_x \left( \hat{x}- \frac{L_x}{2} \right)
         \tanh \left[ \hat{M} \left( \hat{x} - \frac{L_x}{2} \right)
    \right] \; \right\} .
\een

In this work, we consider the flavor non-singlet mesons in QCD with two 
flavors. The hadron correlator is then given by
\begin{eqnarray}
\label{mesonpsop}
 \<H(n)H(0)^{\dagger}\> 
&=& \< G  \> \\ \nonumber
&=&\<\Tr 
\left[ P(\hat{\mu}_u)_{n 0}\Gamma P(\hat{\mu}_d)_{0 n} \Gamma^{\dagger}\right] \> \ . 
\end{eqnarray}
Here $P(\hat{\mu})=D[U; \hat{\mu}]^{-1}$ 
($D[U;\hat{\mu}]$ is the Dirac operator)
is the quark propagator at finite chemical
potential, and $\Gamma$ is the Dirac matrix which
specifies the spin of the meson.
In this paper we study
the response to the isoscalar and isovector chemical potentials 
$\hat{\mu}_S, \hat{\mu}_V$ by setting
\begin{equation}
\hat{\mu}_S = \hat{\mu}_u = \hat{\mu}_d  \hspace{2mm};\hspace{2mm} 
\hat{\mu}_V=\hat{\mu}_u=-\hat{\mu}_d \ .
\label{scal}
\end{equation}

\vspace{-3mm}
\section{Numerical Simulations and Results}
\vspace{-3mm}
The simulations have been performed at finite temperature 
$T/T_c \in \sim[0.9,1.1]$ on a $16 \times 8^2 \times 4$ lattice
with standard Wilson gauge action and with two dynamical flavors of 
staggered quarks. We use the R-algorithm, with quark masses
$ma = 0.0125$, $0.017$ and $0.025$. 
We also use a corner-type wall source after Coulomb gauge 
fixing in each $(y,z,t)$-hyperplane.

The first derivative of the pseudoscalar meson correlator 
with respect to the isoscalar chemical 
potential is identically zero.  For the isovector chemical potential, 
our simulation values for the first derivative are very small in both phases. 

\subsection{Response of the pseudoscalar meson to the 
isoscalar chemical potential}
\begin{figure}[htbp]
\centerline{\psfig{figure=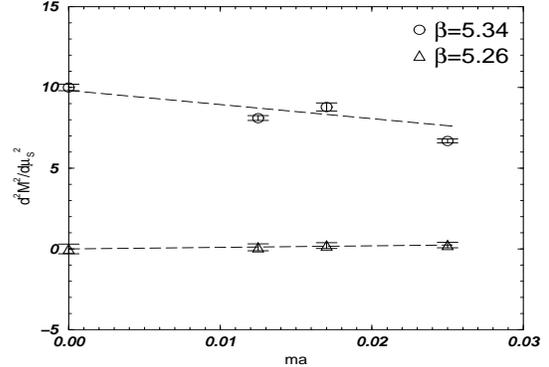,width=7.0cm,height=5cm}}
\vspace{-1cm}
\caption{
$d^2\hat{M}^2/d\hat{\mu}_S^2$  for
the pseudoscalar meson versus $ma$ at $T<T_c$ ($\beta=5.26$, triangles) and 
$T>T_c$ ($\beta=5.34$, circles).
Extrapolation to $ma=0$ is also shown.}
\label{chlimddotM}
\end{figure}

In the low temperature phase, the dependence of the mass on $\hat \mu_S$  
is small. This behavior is to be expected, since, below the critical 
temperature and in the vicinity of zero $\hat \mu_S$, the pseudoscalar  
meson is still a Goldstone boson.
In fact, the chiral extrapolation of the isoscalar
response is consistent with zero, as shown in Figure~\ref{chlimddotM}.
This is in contrast with the behavior above $T_c$, where
$d^2 \hat{M}^2/d \hat{\mu}^2$ seems to remain finite
even in the chiral limit.
In addition, our results suggest that the response of the coupling
$A$ (eq.(\ref{singlepole})) is small below $T_c$.

Above $T_c$, we first note that the correlator and its response
are still well fitted by the single pole formulae,
Eqs.~(\ref{singlepole}-\ref{deri-2}).
The screening masses are manifestly larger than those below $T_c$.
As pointed out above, the response of the mass above $T_c$ becomes large and
 positive, reflecting the fact that the pion is no longer a Goldstone
boson and indicating chiral symmetry restoration.

\subsection{Response of the pseudoscalar meson to the
isovector chemical potential}
In the presence of the isovector chemical potential, $\pi^{+}$ and
$\pi^{-}$ may have different masses. Here we consider the $\pi^{+}$
 ( $u \bar{d}$ ) meson  as in Eq.~(\ref{mesonpsop}).

\begin{figure}[h]
\centerline{\psfig{figure=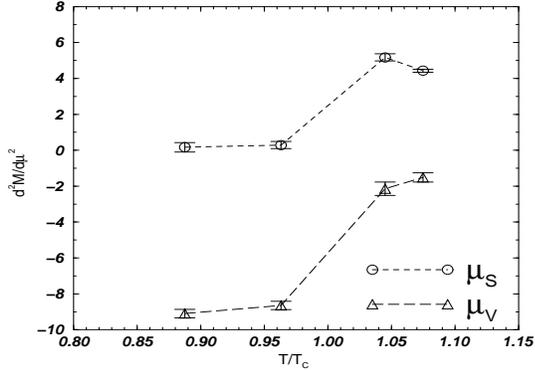,width=7.0cm,height=5cm}}
\vspace*{-1cm}
\caption{Second responses $d^2\hat{M}/d\hat{\mu}^2_S$ and
$d^2\hat{M}/d\hat{\mu}^2_V$ of the pseudoscalar meson mass at $ma=0.025$.
}
\label{fig:ddotM025}
\end{figure}

An interesting point in this respect is that the second derivative of the mass 
is negative in the low temperature phase, in contrast with the isoscalar 
potential case: the mass tends to 
decrease under the influence of the isovector chemical potential. 
This may be explained by the observation 
that, for low temperature and chemical potential
above the pion mass,
a Goldstone mode can appear~ \cite{Stephanov}.
The behavior of $M$ is more clearly shown by a  Taylor expansion:  
At $\beta=5.26$ and $ma=0.017$, the
data suggest

\vspace{2mm}
$\left. \frac{M(\mu_V)}{T}\right|_{\mu_V}
=  1.4024 (8)
-0.0005(10) \left( \frac{\mu_V}{T} \right)
\\ 
-1.31(4) \left( \frac{\mu_V}{T} \right)^2
+ O \left[ \left(\frac{\mu_V}{T}\right)^3 \right].
$
\vspace{2mm}

In the high temperature phase, the dependence of the masses on $\mu_V$
decreases. Since the pseudoscalar meson becomes heavier, the phase boundary
to the pion condensate phase is further away from the $ \mu_V=0$ axis.
The weaker responses may be understood from this point of view.
\begin{figure}[htbp]
\centerline{\psfig{figure=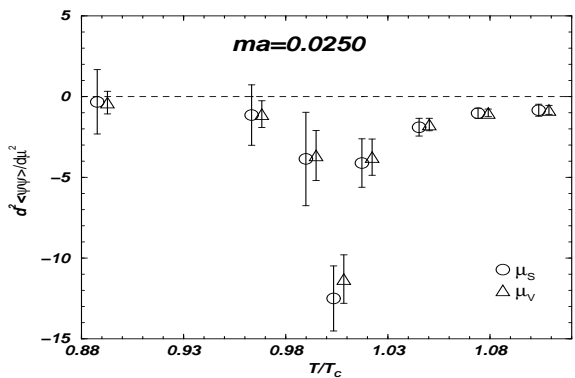,width=7.5cm,height=5cm}}
\vspace*{-1cm}
\caption{Second responses $d^2{\< \chiral \>}/d\hat{\mu}^2_S$ and
$d^2{\< \chiral \>}/d\hat{\mu}^2_V$  at $ma=0.025$.
Note that the abscissa of coincident points have been splitted.
}
\label{fig:psi}
\end{figure}
\subsection{Responses of the chiral condensate }
We also measure responses of the chiral condensate
to the isoscalar and isovector chemical potentials. 
The first responses to both potentials are identically zero. 
Figure~\ref{fig:psi} shows our preliminary results for the second responses.  
We have found they are small and negative  in both phases. 
Near $T_c$ there is a large change. The critical temperature
tends to decrease under the influence of $\mu_{S,V}$. 
Thus, in the low temperature phase, turning on the chemical potential brings 
the system closer to the phase transition where chiral symmetry is restored,
and decreases the chiral condensate.
At high temperature, because chiral symmetry is restored, 
responses of the chiral condensate to the isoscalar 
and isovector chemical potential 
are small. 
\vspace{-0.30cm}
\section{Conclusions}
\vspace{-0.20cm}
We have developed a framework to study the response of hadrons 
to the chemical potential.  
The dependence 
of the pseudoscalar meson mass on $ \mu_S$  in the chiral limit is
consistent with zero at low temperature, reflecting the fact that at 
small $ \mu_S$ the pion  is still a Goldstone boson.
For the isovector chemical potential, 
the $u \bar{d}$ pseudoscalar meson mass tends to decrease as a function
of $ \mu_V$ at a much stronger rate in the low temperature phase.
This is consistent with a restoration of chiral symmetry as the
isovector chemical potential increases.


\end{document}